\documentclass[review]{elsarticle}
\usepackage{amsmath}
\usepackage{lineno,hyperref}
\modulolinenumbers[205]
\journal{Journal of \LaTeX\ Templates}
\usepackage{comment}
\usepackage{setspace}
\usepackage{hyperref}
 
\usepackage{geometry}
\usepackage{color}
\usepackage[normalem]{ulem}
\newgeometry{vmargin={21mm}, hmargin={20mm,20mm}}

\begin{document}

\begin{frontmatter}
\title{The generation of diverse traveling pulses and its solution scheme in an excitable slow-fast dynamics}
\tnotetext[]{Corresponding author\\ Email address: arghamondalb1@gmail.com}
\author{Arnab Mondal$^{1}$, Argha Mondal$^{2,3,*}$, M. A. Aziz-Alaoui$^{4}$, Ranjit Kumar Upadhyay$^{1}$,  Sanjeev Kumar Sharma$^1$, Chris G. Antonopoulos$^3$}
\address{$^{1}$ Department of Mathematics \& Computing, Indian Institute of Technology (Indian School of Mines), Dhanbad 826004, India\\
	$^2$ Department of Mathematics, Sidho-Kanho-Birsha University, Purulia 723104, WB, India\\
	$^3$ Department of Mathematical Sciences, University of Essex, Wivenhoe Park CO4 3SQ, UK\\
	$^4$ Normandie Univ, UNIHAVRE, LMAH, FR-CNRS-3335, ISCN, 76600 Le Havre, France
}

%
%
%
%

\begin{abstract}
In this paper, we report on the generation and propagation of traveling pulses in a homogeneous network of diffusively coupled, excitable, slow-fast dynamical neurons. The spatially extended system is modelled using the nearest neighbor coupling theory, in which the diffusion part measures the spatial distribution of the coupling topology. We derive analytically the conditions for traveling wave profiles that allow the construction of the shape of traveling nerve impulses. The analytical and numerical results are used to explore the nature of the propagating pulses. The symmetric or asymmetric nature of the traveling pulses is characterized and the wave velocity is derived as a function of system parameters. Moreover, we present our results for an extended excitable medium by considering a slow-fast biophysical model with a homogeneous, diffusive coupling that can exhibit various traveling pulses. The appearance of series of pulses is an interesting phenomenon from biophysical and dynamical perspective. Varying the perturbation and coupling parameters, we observe the propagation of activities with various amplitude modulations and transition phases of different wave profiles that affect the speed of the pulses in certain parameter regimes. We observe different types of traveling pulses, such as envelope solitons and multi-bump solutions and show how system parameters and the coupling play a major role in the formation of different traveling pulses. Finally, we obtain the conditions for stable and unstable plane waves.
\end{abstract}

 
\end{frontmatter}

\begin{quotation}
\noindent\textbf{The emergence of diverse traveling wave profiles might reflect the rhythmic properties in signal processing and functional connectivity in different brain areas. In \cite{mondal2018dynamics}, we have shown the existence of solitary wave profiles in the Hindmarsh-Rose model using the tanh method. Here, we use a different analytical approach to find the traveling wave profiles for a slow-fast FitzHugh-Rinzel model. We observe different types of traveling pulses, such as envelope solitons and multi-bump solutions and show how system parameters and the coupling play a major role in the formation of various traveling pulses. Finally, we obtain the conditions for which plane waves become stable or unstable.}
 \end{quotation}

\section{Introduction}\label{sec:intro}

Excitable biophysical mechanisms show various dynamical characteristics and can be studied from a dynamical systems perspective. Neurons receive incoming sensory inputs, encode them into various biophysical variables and produce relevant outputs \cite{izhikevich2001synchronization,izhikevich2007dynamical}. Theoretical modeling and numerical simulations are prominent tools in understanding various functional mechanisms in neural computations \cite{ma2017review,iqbal2017pattern,muruganandam1997bifurcation}.  Hindmarsh-Rose (HR) \cite{corson2009asymptotic}, Izhikevich \cite{izhikevich2001synchronization,izhikevich2007dynamical}, FitzHugh-Nagumo (FHN) \cite{fitzhugh1961impulses} and FitzHugh-Rinzel (FHR) \cite{rinzel1986nonlinear,rinzel1987mathematical} systems are such types of biophysical models that produce diverse firing properties. In extended spatial systems of coupled neuronal populations, a transmembrane potential difference can travel across neurons by means of traveling wave propagation \cite{townsend2018detection,milton1993spiral,kondo2010reaction,meier2015bursting,feng2016spike}. This may cause the appearance and disappearance of neuronal responses with traveling pulses. Such types of dynamics are responsible for playing a major role in the biodynamics of signal processing \cite{belykh2005synchronization}.

Traveling pulses present a characteristic feature of certain neurological disorders in humans, including epilepsy \cite{connors1993generation} and migraines \cite{lance1993current}. Therefore, investigating the mechanisms underlying wave propagation in neuronal tissue is important in understanding both normal and pathological brain states. The spatial propagation of wavy profiles in coupled neuronal networks can emerge due to variations in the electrical activities at the single cell level. One of the key challenges is to analyze the intrinsic dynamics of neuronal activities and its characteristics \cite{connors1990intrinsic,izhikevich2004model,yafia2013existence}. The emergence of diverse traveling wave profiles is often associated with self-organization and interactions within the network and might reflect the rhythmic properties in signal processing, and functional connectivity in different brain areas \cite{wig2011concepts,folias2005stimulus,kilpatrick2010effects}. 

The main goal here is to study the emergence of diverse traveling pulses using a mathematical approach for spatially excitable media in the context of excitable biophysical systems. We combine the existence and propagation of traveling pulses with speed equations and the stability of the solutions. Traveling pulses are special types of solutions to a reaction diffusion system. Equilibrium solutions can lead to two types of traveling waves known as traveling fronts and pulse solutions. A traveling front connects distinct steady states at both ends, however, a traveling pulse connects the same steady state \cite{pinto2001spatially,kilpatrick2008traveling}.

Our study is motivated by theoretical and experimental works that assume the existence of localized excitations in certain neuronal populations \cite{ermakova2009propagation,villacorta2013wave,kakmeni2014localized,tankou2019localized,mondal2018dynamics}. Many research articles have investigated the existence of envelope solitons of the nerve impulse, not only in diffusively connected networks but also in memristive neuron models \cite{kakmeni2014localized,tankou2019localized}. In \cite{mondal2018dynamics}, we have shown the existence of solitary wave profiles in the HR model using the {\it tanh} method \cite{malfliet1996tanh,wazwaz2005tanh}. Further, a special type of traveling pulse, i.e., the multi-bump or multi-pulse solution has been studied in \cite{kilpatrick2008traveling}. Specific mechanisms for traveling pulses have been examined mathematically. Excitable neuron models can produce two types of traveling pulses: envelope solitons and envelope of multi-bump solutions. Envelope solitons are observed in nonlinear and weakly dispersive systems. The wave equation in the small amplitude limit is very important to describe envelope solitons. The disappearance of bumps changes the propagation activity. The individual bump solution in multi-pulse wave profiles are non-propagating and transient \cite{kilpatrick2008traveling}. However, there arises important questions about the existence of analytical, exact or approximate solutions to the excitable models as it is interesting to test the numerical solutions with theoretical approaches.

Here, we investigate different forms of nerve impulses in terms of traveling wave profiles in a diffusively coupled network of slow-fast spiking-bursting FHR type 3 neuron models. The main objectives are analytical and numerical approaches to find the dynamical behavior of modulated traveling pulses in an excitable diffusive network of slow-fast dynamics. We use the multiple-scale expansion method in a semi-discrete approximation to obtain the modified complex Ginzburg-Landau (CGL) equation using perturbation theory. Next, we derive the envelope of the traveling pulses to observe the collective dynamics, i.e., nerve impulses. Finally, we derive the role of this type of connected oscillators with identical coupling and the effects of small perturbations on the solution of the wave profiles to observe both the envelope of multipulses and single solitary pulses.

The study of the propagation of various traveling pulses in such networks may be relevant, e.g., in brain pathologies \cite{stefanescu2008low} and different functional mechanisms especially in cortical areas \cite{kakmeni2014localized}. Our work describes an explicit analytical scheme and numerical approaches to analyze the diffusively coupled FHR models for a range of coupling strengths. First the network can be considered in the Lienard form and then it can be transformed into an extended complex Ginzburg-Landau equation (CGLE) using perturbation theory. We apply the multiple-scale expansion theory with a semi-discrete approximation. The reduced CGLE follows the time evolution of the modulated wave propagation in the network. To obtain the traveling wave solution, we derive the expression of the nerve impulse using the membrane voltage variable.

The paper is organized as follows. In Sec. \ref{sec2}, we describe preliminaries about the slow-fast subsystems and present the FHR type 3 neuron model. We demonstrate the theory to derive the approximate solution using semi-discrete approximations. In Sec. \ref{sec3}, we reduce the FHR model to the CGLE and find in Sec. \ref{sec4}, the solution to the latter. We also present different types of traveling pulses, such as envelope solitons and multi-bump solutions. Further, we discuss how these depend on system parameters. In Sec. \ref{sec5}, we obtain the conditions for which the plane wave becomes stable or unstable. Finally, we conclude our results in Sec. \ref{sec6} following a brief discussion.

\section{The biophysical excitable model and the approach to finding its solution}\label{sec2}

\subsection{Formulation of the coupled network model}

The dynamics of the minimal mathematical model to explore different time scales of slow and fast membrane voltage processes can be defined as $dx/dt = F (x, z)$, $dz/dt = \mu G(x, z)$, where $x \in R^2$ and $z \in R$ with $0 < \mu \ll 1$. The fast and slow subsystems are 2D and 1D abstractions of the dynamics while facilitating the geometrical features in the phase space. The dynamics of the system evolves slowly around the slow variable when bursting interacts dynamically with the fast subsystem. The transition between quiescent and firing states can be studied by performing a bifurcation analysis where the bifurcation parameter is $\mu$. The slow-fast model can be presented as a generic model of excitabilities with diverse neuronal responses. It exhibits different nonlinear properties that are governed by the set of three coupled ODEs in Eq. \eqref{model}. Our primary goal is to study the dynamics on a spatially extended medium with system  \eqref{model} exhibiting various oscillatory firing patterns.

The FHR model was introduced by FitzHugh and Rinzel \cite{rinzel1986nonlinear,rinzel1987mathematical,rinzel1982bursting,wojcik2015voltage,mondal2019firing}, and is an extended modification of the 2D classical FHN model \cite{fitzhugh1961impulses}. A $z$-shaped slow manifold can be observed in the case of the traditional bursting model, while the FHR model of type 3 bursting does not have slow manifolds \cite{nergo1998evidence}. The most important variable is the membrane voltage $v$. The third variable of the model enables us to understand the control of the resting period between the generation of two action potentials. The equations of an excitable network of $N$ neurons with gap junction coupling are given by
\begin{align}
\dot{v}_k (t) &= w_k -4(v_k^3 -v_k) - z_k +I \nonumber\\&+D(v_{k+1} -2v_k +v_{k-1}),\nonumber\\
\dot{w}_k (t) &= - (4v_k +1 +w_k),\label{model}\\
\dot{z}_k (t) &= \mu (1.25v_k - (z_k - z_0)/4)\nonumber,
\end{align}
where $k = 1,\ldots, N$, $v$ is the membrane voltage and $w$ the recovery variable. $I$ measures the magnitude of the external input stimulus current. $z$ is a slow variable, and $z_0$ the slow subsystem parameter that controls the nature of bursting. The small parameter $0<\mu\ll1$ determines the pace of the slow variable $z$ and $D$ is the coupling strength of the gap junctions.

We have already studied the dynamics of system \eqref{model} in \cite{mondal2021spatiotemporal}. In this work, we consider the case of two nearest-neighbors coupled with weak coupling. Neurons can connect with their nearest neighbors electrically. The weak coupling reflects the situation that appears in the study of bursting-like activity in $\beta$-cell islets of the pancreas that secrete insulin \cite{kakmeni2014localized,perez1991biophysical,raghavachari1999waves}. Biophysically, the neuronal variability may reflect the characteristics at different levels of certain receptors or differences in regulatory effects that can be induced by internal or external neuron-modulatory processes.

In the following, we will use the semi-discrete approximation \cite{tankou2019localized,kakmeni2014localized} to derive analytically the localized traveling pulses that propagate in a diffusively connected excitable network of FHR neurons \eqref{model}. In order to investigate the resulting interactions, the system is considered in such a way that each neuronal node follows the same coupling topology. The dynamics of spike generation can be described by the dynamics of the FHR model \eqref{model}.

\subsection{The solution method}

Large scale coupled neuronal networks can generate various spatial states with different structures. One of these states give rise to traveling pulses. Here we derive analytically the emergence of traveling pulses using the semi-discrete approach \cite{tankou2019localized,kakmeni2014localized}. In particular, we transform system \eqref{model} in the wave form by reducing its first and second equations into the second order differential equation
\begin{align}
\ddot v_k + f(v_k, \dot v_k, z_k) &= D(v_{k+1} -2v_k +v_{k-1}) \nonumber\\&+ D(\dot v_{k+1} - 2\dot v_k + \dot v_{k-1})\label{lienard},\\
\dot z_k &= g(v_k, z_k),\nonumber
\end{align}
where the first equation is a second-order ordinary differential equation. Equation \eqref{lienard} is also known as the Lienard form. The associated constant parameters are to be derived during the study of the semi-discrete approximation. We note that this transformation does not affect the behavior of the system. We can easily decompose the system and obtain the transformed slow-fast system. 

To study different traveling wave profiles, we treat the slow-fast system as a discrete coupled system. The evolution of modulated waves in the network is described by a modified CGLE. To find the traveling wave profiles, we need to reduce the system into a CGLE and find its solution using the semi-discrete approximation \cite{tankou2019localized,kakmeni2014localized}. We consider the new variables $m_k$ and $n_k$ as
\begin{align}
v_k&=\epsilon m_k\\ 
z_k&=\epsilon n_k,
\end{align}
where $0<\epsilon \ll 1$. Then, system \eqref{lienard} becomes
\begin{align}
\ddot m_k + f_1(m_k, \dot m_k, n_k) &= D(m_{k+1} -2m_k +m_{k-1}) \nonumber\\&+ D(\dot m_{k+1} - 2\dot m_k + \dot m_{k-1}),\label{lienard2}\\
\dot n_k &= g_1(m_k, n_k).\nonumber
\end{align}

According to the method, which is a perturbation technique, the carrier waves are kept discrete whereas the amplitude is considered in the continuum limit \cite{kakmeni2014localized, tankou2019localized}. The semi-discrete approximation approach is used to study plane wave modulation due to the nonlinear terms in the system. The main objective to use the multiple-scale expansion method is to find the solution $v(x,t)$ in terms of the new independent space and time variables $X_k$ and $T_k$, that we discuss next. Applying the approximation, the method allows one to study the modulation of a plane wave due to nonlinear effects \cite{kakmeni2014localized}.

In this context, we introduce the new space and time variables $X_k=\epsilon^k x$ and $T_k=\epsilon^k t$, where $x$ and $t$ are the space and time variables. Then, we consider
\begin{align*}
v(x,t)&=\sum_{k=1}^{\infty}\epsilon^k m_k (X_0,X_1,X_2,\ldots T_0,T_1,T_2,\ldots),\\
z(x,t)&=\sum_{k=1}^{\infty}\epsilon^k n_k (X_0,X_1,X_2,\ldots T_0,T_1,T_2,\ldots).
\end{align*}
The partial derivatives with respect to the new space and time variables are given by
\begin{align*}
\frac{\partial}{\partial t}&=\frac{\partial}{\partial T_0} + \epsilon \frac{\partial}{\partial T_1} + \epsilon^2 \frac{\partial}{\partial T_2} + \ldots\\
\frac{\partial}{\partial x}&=\frac{\partial}{\partial X_0} + \epsilon \frac{\partial}{\partial X_1} + \epsilon^2 \frac{\partial}{\partial X_2} + \ldots.
\end{align*}
Finally, to obtain the solution, we use the new form of the voltage and slow variables, the time and space derivatives and compare various terms at different orders of $\epsilon$.

\section{Equation of motion of the amplitude (CGLE)} \label{sec3}
To study the envelope solitons of wave equations in the small amplitude limit, we find the solution to this equation analytically and compute it numerically. To proceed, it is important to transform system \eqref{model} into the wave form. To do so, we introduce a second-order differential equation in $v_k$ for the first and second equations in model \eqref{model} and obtain the system of ODEs
\begin{align}
\ddot v_k +f(v_k, \dot v_k, z_k) &= D(v_{k+1} -2v_k +v_{k-1})\nonumber \\&+D_1(\dot v_{k+1} - 2\dot v_k + \dot v_{k-1}),\nonumber\\
\dot z_k &= g(v_k, z_k),\label{model2D}
\end{align}
where $f(v_k, \dot v_k, z_k)=E_0 v_k + (E_1+E_2v_k^2)\dot v_k + \frac{{{E_2}}}{3}v_k^3 + E_3 z_k + I_0$, $g(v_k, z_k)=\mu ( 1.25v_k - (z_k - z_0)/4)$, $E_0=1.25 \mu$, $E_1=-3$, $E_2=12,$ $E_3=1-\frac{\mu}{4}$, $I_0=1+\frac{\mu z_0}{4} - I$ and $D_1=D$. This transformation does not change the system \eqref{model}, it only reduces its dimension from 3 to 2. The transformed system \eqref{model2D} is equivalent to system \eqref{model} and we can obtain approximate solutions of the nonlinear system \eqref{model2D} using a perturbation technique. As we need to derive the solution in an extended diffusive medium, we will use the perturbation $v_k=\epsilon m_k$, $z_k=\epsilon n_k$ and perturb the parameters $E_1$, $E_3$ and $D_1$ up to order $\epsilon^2$. Then, system \eqref{model2D} becomes
\begin{eqnarray}
\ddot m_k + f_1(m_k, \dot m_k, n_k) &=& D(m_{k+1} -2m_k +m_{k-1}) \label{model2D1}\\&+& \epsilon^2 D_1(\dot m_{k+1} - 2\dot m_k + \dot m_{k-1}),\nonumber\\
\dot n_k &=&g_1(m_k, n_k),\label{model2D2}
\end{eqnarray}
where $f_1(m_k, \dot m_k, n_k)=E_0 m_k + \epsilon^2(E_1+E_2m_k^2)\dot m_k + \epsilon^2\frac{E_2}{3}m_k^3 +\epsilon^2 E_3 n_k$, $g_1(m_k, n_k)=-sn_k + E_0 m_k$ and $s=\frac{\mu}{4}$. Equation \eqref{model2D1} generates the dynamical behavior of the membrane voltages in coupled FHR neurons. Equation \eqref{model2D2} reflects the associated coupling term that describes the dynamics of the bursting variable. Since we are dealing with a weakly coupled diffusively connected network with nonlinear excitations, it is possible to use the semi-discrete approximation. In the following section, we will explore analytically and numerically the traveling pulses as envelope solitons. We will demonstrate the method to obtain the modified CGLE. Next, we reduce it to a CGLE by using multiple-scale expansions to find its solution.

We consider the solution of Eqs. \eqref{model2D1}-\eqref{model2D2} in the following forms
\begin{align}
m_k&=A_k e^ {i\theta_k} + \overline{A}_k e^ {-i\theta_k} + \epsilon \Big(B_k + C_k e^ {2i\theta_k} + \overline{C}_k e^ {-2i\theta_k}\Big),\label{model2Dtransformation}\\
n_k&=F_k e^ {i\theta_k} + \overline{F}_k e^ {-i\theta_k} + \epsilon (G_k + H_k e^ {2i\theta_k} + \overline{H}_k e^ {-2i\theta_k})\nonumber,
\end{align}
where $\theta_k=qk-\omega t$. $q$ and $\omega$ indicate the normal mode wave vector and angular velocity, respectively. Here, we use the continuum limit approximation on the amplitudes $A_k(t)$, $B_k(t)$ and $C_k(t)$ ($F_k(t)$, $G_k(t)$ and $H_k(t)$, respectively) as they change slowly with respect to space and time. Using the continuum limit approximation \cite{kakmeni2014localized} on the amplitudes $A_k(t)$, $B_k(t)$ and $C_k(t)$, the amplitudes become $A(X_1,X_2,T_1,T_2)$, $B(X_1,X_2,T_1,T_2)$ and $C(X_1,X_2,T_1,T_2)$ (similarly for $F$, $G$ and $H$). Using Taylor expansions, $A_{k \pm 1}$ is given by
\begin{equation}
A_{k \pm 1}=A \pm \epsilon \frac{\partial A}{\partial X_1} \pm \epsilon^2 \frac{\partial A}{\partial X_2} + \frac{\epsilon^2}{2} \frac{\partial^2 A}{\partial X_1^2} + O(\epsilon^3).
\end{equation}
The first and second-order temporal derivatives are given by
\begin{equation}
\frac{\partial A_k}{\partial t}=\epsilon \frac{\partial A}{\partial T_1} + \epsilon^2 \frac{\partial A}{\partial T_2} + o(\epsilon^3)
\end{equation}
and 
\begin{equation}
\frac{\partial^2 A_k}{\partial t^2}=\epsilon^2 \frac{\partial^2 A}{\partial T_1^2} + o(\epsilon^3),
\end{equation}
respectively. Similar expressions hold for the amplitudes $B$, $C$, $F$, $G$ and $H$. The first and second-order temporal derivatives of $m_k$ with respect to the new space and time variables and amplitudes $A$, $B$, $C$ are given by
\begin{align*}
\dot m_k&= \left(\epsilon \frac{\partial A}{\partial T_1} + \epsilon^2\frac{\partial A}{\partial T_2} - i\omega A\right)e^{i\theta_k} \\& + \left(\epsilon \frac{\partial \bar{A}}{\partial T_1} + \epsilon^2\frac{\partial \bar{A}}{\partial T_2} + i\omega \bar{A}\right)e^{-i\theta_k} +\epsilon^2\frac{\partial B}{\partial T_1} \\&+ \left(\epsilon^2\frac{\partial C}{\partial T_1} - \epsilon 2i\omega C\right)e^{2i\theta_k} \\&+\left(\epsilon^2\frac{\partial \bar{C}}{\partial T_1} + \epsilon 2i\omega \bar{C}\right) e^{-2i\theta_k} + o(\epsilon^3),\\
\ddot m_k&= \left(\epsilon^2 \frac{\partial^2 A}{\partial T_1^2} - \epsilon 2i\omega \frac{\partial A}{\partial T_1} - \epsilon^2 2i\omega \frac{\partial A}{\partial T_2} - \omega^2 A\right)e^{i\theta_k}\\& + \left(\epsilon^2 \frac{\partial^2 \bar{A}}{\partial T_1^2} + \epsilon 2i\omega \frac{\partial \bar{A}}{\partial T_1} + \epsilon^2 2i\omega \frac{\partial \bar{A}}{\partial T_2} - \omega^2 \bar{A}\right)e^{-i\theta_k} \\&+\left(-\epsilon^2 4i\omega \frac{\partial C}{\partial T_1} - \epsilon 4\omega^2 C\right)e^{2i\theta_k} \\&+ \left(\epsilon^2 4i\omega \frac{\partial \bar{C}}{\partial T_1} - \epsilon 4\omega^2 \bar{C}\right)e^{-2i\theta_k} + o(\epsilon^3).
\end{align*}
Similarly, the first order temporal derivatives of $n_k$ with respect to the new space and time variables and amplitudes $F$, $G$, $H$ are given by 
\begin{align*}
\dot n_k&= \left(\epsilon \frac{\partial F}{\partial T_1} + \epsilon^2\frac{\partial F}{\partial T_2} - i\omega F\right)e^{i\theta_k} \\&+ \left(\epsilon \frac{\partial \bar{F}}{\partial T_1} + \epsilon^2\frac{\partial \bar{F}}{\partial T_2} + i\omega \bar{F}\right) e^{-i\theta_k} \\&+\epsilon^2\frac{\partial G}{\partial T_1} + \left(\epsilon^2\frac{\partial H}{\partial T_1} - \epsilon 2i\omega H\right)e^{2i\theta_k}\\&+ \left(\epsilon^2\frac{\partial \bar{H}}{\partial T_1} + \epsilon 2i\omega \bar{H}\right)e^{-2i\theta_k} + o(\epsilon^3).
\end{align*}
Substituting the above equations in Eqs. \eqref{model2D1} - \eqref{model2D2} (see Appendix A) and equating the coefficients of different orders in $\epsilon$, we obtain the following important relations:

The dispersion relation of linear pulses by comparing the coefficient of $e^{\pm i \theta_k}$ in Eq. \eqref{model2D1} (see Eq. (A1)) is given by
\begin{equation}
\omega^2=E_0+4D\sin^2 \left(\frac{q}{2}\right).
\end{equation}

\begin{figure*}[!ht]
\centering
\includegraphics[width=14cm,height=5cm]{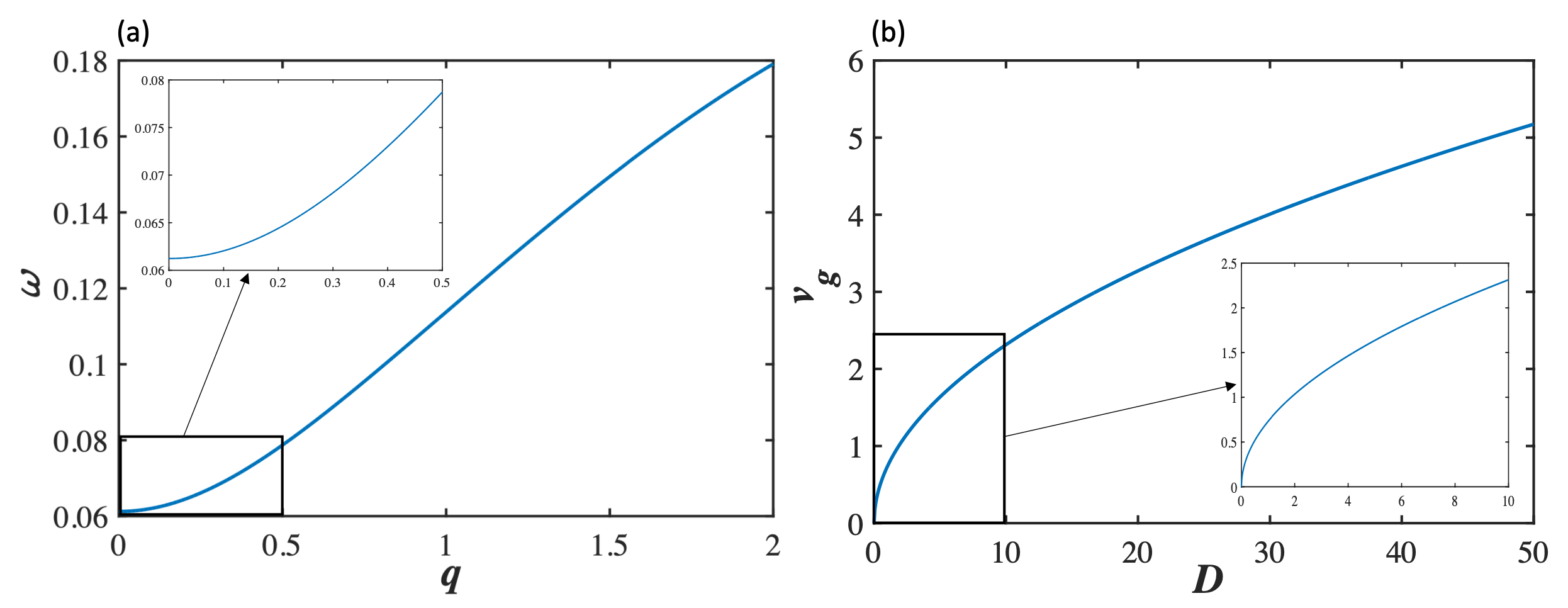}
\caption{The effect of $q$ and $D$ on $\omega$ and $v_g$. (a) Dispersion relation of linear pulses for $\mu=0.003$ and $D=0.01$. (b) The effect of the diffusion coefficient $D$ on the velocity of the pulses for $\mu=0.003$, $q=1.5$. At smaller diffusion coefficient, velocity $v_g$ changes very rapidly, whereas for larger diffusion coefficient, the rate of change of velocity is very low. The insets show a zoom-in of the behavior in the black rectangles.}\label{fig1}
\end{figure*}
We have also shown the dispersion relation numerically in Fig. \eqref{fig1}(a). The dispersion relation of the traveling pulses in the network depends on the associated parameters of the coupled system. 

Similarly, comparing the coefficient of $\epsilon e^{i \theta_k}$ in Eq. \eqref{model2D1} (see Eq. (A1)), we obtain
\begin{equation}
\frac{\partial A}{\partial T_1} + v_g \frac{\partial A}{\partial X_1}=0,
\end{equation}
where $v_g$ is known as the group velocity, given by
\begin{equation}
v_g=\frac{D \sin q}{\omega}.
\end{equation}
Clearly, the diffusion coefficient $D$ plays a major role in controlling the velocity of the  pulses. We have shown the effects of diffusive coupling on the velocity of the pulses (Fig. \ref{fig1}(b)). Interestingly, for smaller diffusion coefficient, the velocity $v_g$ changes very rapidly, whereas for higher diffusion coefficient, the rate of change of the velocity is very low. We have shown a zoomed version in Fig. \ref{fig1}(b). The dispersion equation is related to the parameters associated with the excitable network. The wave velocity depends on the diffusive property of the membrane dynamics, as the nerve impulses move faster according to the ionic movements in the cell membrane. 

Equating the coefficients without exponential terms and $\epsilon e^{2i\theta_k}$ in Eq. \eqref{model2D1}, we obtain $B=0$ and $C=0$, respectively. As a result of the coupling between the membrane voltage and slow variables, we get 
\begin{equation*}
F=\frac{E_0(s+i  \omega)}{s^2+\omega^2}A
\end{equation*}
by equating the coefficients of $e^{i\theta_k}$ in Eq. \eqref{model2D2} (see Eq. (A2)). 

The terms depending on $\epsilon^2 e^{i\theta_k}$ have the following relation in Eq. \eqref{model2D1} (see Eq. (A1))
\begin{align}
\frac{\partial^2 A}{\partial T_1^2} - 2i\omega \frac{\partial A}{\partial T_2}&=i\omega E_1A\nonumber \\&+(i\omega -1)E_2\mid A \mid^2A - \frac{E_0E_3(s+iw)}{s^2+w^2}A \nonumber\\&+ 4i\omega D_1 \sin^2 (\frac{q}{2})A + 2iD\sin (q)\frac{\partial A}{\partial X_2}\nonumber \\&+D\cos (q)\frac{\partial^2 A}{\partial X_1^2}.
\end{align}
Using the transformation $u_k = X_k - v_gT_k$ and $ \tau_k=T_k$, we obtain
\begin{equation}
i\frac{\partial A}{\partial \tau_2} + \frac{P}{2} \frac{\partial^2 A}{\partial u_1^2} +Q \mid A \mid^2A + i \frac{R}{2} A =0.\label{CGLE}
\end{equation}
Equation \eqref{CGLE} is the modified CGLE, which describes the evolution of modulated pulses in the neuronal network.

The real dispersion coefficient $P$ is given by 
\begin{equation*}
P=\frac{D \omega ^2\cos q - D^2\sin^2q}{\omega ^3}.
\end{equation*}

The two complex dissipation coefficients $Q$ and $R$ are given by 
\begin{align*}
Q&=-\frac{E_2}{2\omega} + i\frac{E_2}{2},\\
R&=E_1+4D_1\sin^2\frac{q}{2}-\frac{E_3 E_0}{s^2 + \omega ^2} +i\left (\frac{sE_3 E_0}{w(s^2 +\omega ^2)}\right ).
\end{align*}
Here $Q_r$, $R_r$ and $Q_i$, $R_i$ are the real and imaginary parts of $Q$, $R$ respectively. 

\begin{figure}
\centering
\includegraphics[width=9cm,height=5.5cm]{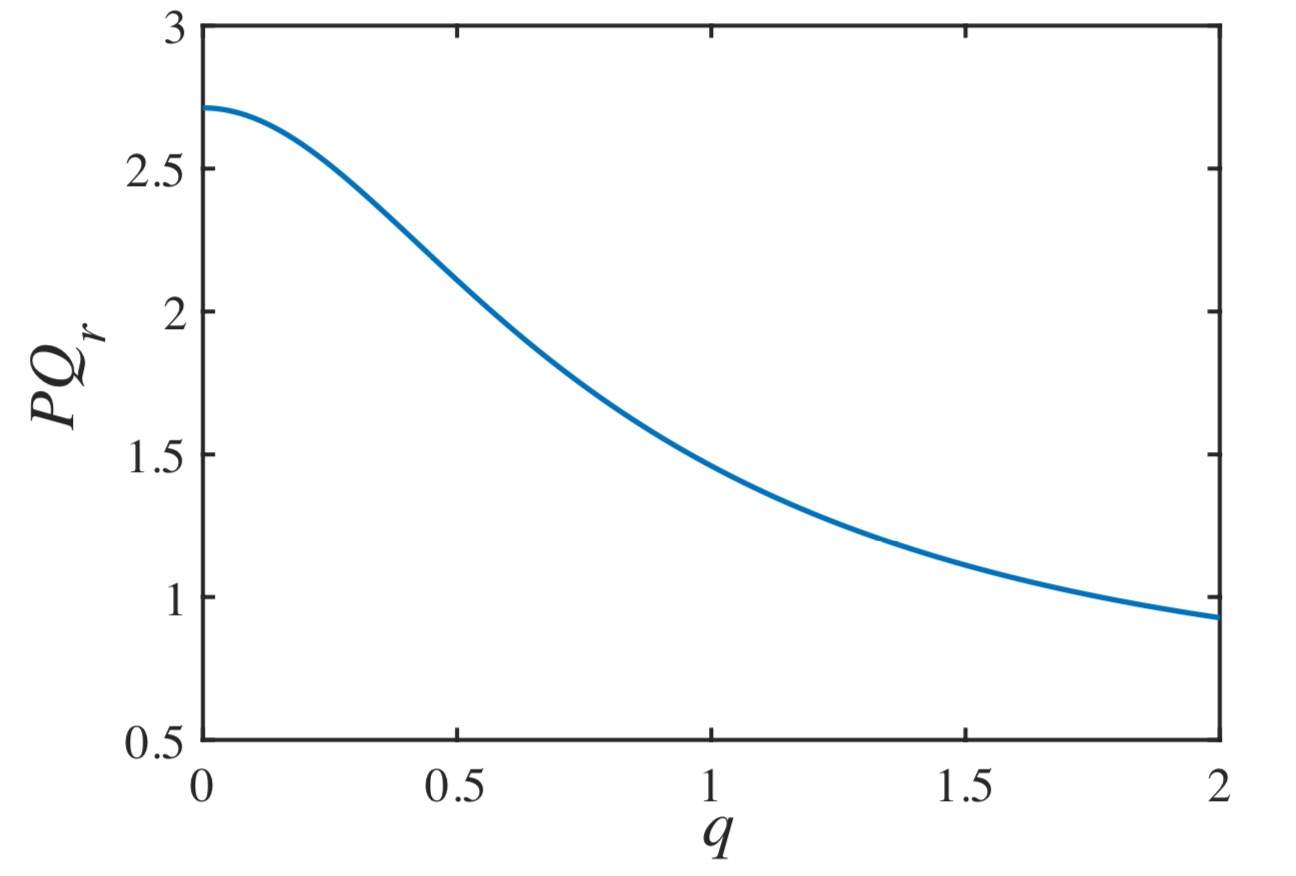}
\caption{Variations of the coefficient $PQ_r$ with respect to the wave vector $q$ for $\mu=0.003$ and $D=0.01$.}\label{fig2}
\end{figure}

The sign of $PQ_r$ plays a major role in determining the modulational instability as the dispersion coefficient is real. According to the Benjamin-Feir instability \cite{tankou2019localized}, positive and negative values of $PQ_r$ indicate that plane pulses are unstable and stable, respectively. Clearly, this stability condition is independent of wave propagation. Thus, in the positive domain of $PQ_r$, we can find the nerve impulses for any wave carrier. We have shown the variations of the coefficient, $PQ_r$ in Fig. \ref{fig2} with respect to the wave vector, $q$.

\section{Traveling wave profiles and numerical results}\label{sec4}

In the following, we obtain the solution of Eq. \eqref{CGLE} to understand the dynamics of the envelope solitons by considering the purely real dissipation term $R$ ($R_i=0$). We consider the solution of Eq. \eqref{CGLE} in the following form
\begin{equation}
A(u_k,\tau_2)=\frac{A_0e^{\alpha}}{1+e^{(\alpha + \bar{\alpha})^(1+i\beta)}},\label{solution}
\end{equation}
where $\alpha=qu_k-\omega \tau_2$, $\beta= \gamma \pm \sqrt{2+\gamma^2}$ and $\gamma=\frac{3Q_r}{2Q_i}$. Substituting Eq. \eqref{solution} into Eq. \eqref{CGLE}, we obtain
\begin{align}
A&=A_0\frac{e^{-\alpha}+\cos 2\alpha \beta e^{\alpha}}{2(\cosh 2\alpha + \cos 2\alpha \beta)}\nonumber \\&+i\left(-A_0\frac{\sin 2\alpha \beta e^{\alpha}}{2(\cosh 2\alpha + \cos 2\alpha \beta)}\right)\label{A}.
\end{align}
From Eq. \eqref{model2Dtransformation}, we obtain
\begin{equation}
m=2(A_r \cos\theta - A_i \sin \theta),\label{Anew}
\end{equation}
where $A_r$ and $A_i$ indicate the real and imaginary parts of $A$. Finally, using Eqs. \eqref{A}, \eqref{Anew} and the expression $v_k=\epsilon m_k$, we obtain
\begin{equation}
v_k=\epsilon A_0 \frac{\cos(\theta_k - 2\alpha_k \beta)e^{\alpha_k} + \cos (\theta_k) e^{-\alpha_k}}{2(\cosh(2 \alpha_k) + \cos (2\alpha_k \beta))}.\label{Eq17}
\end{equation}

\begin{figure*}
\centering
\includegraphics[width=14cm,height=12cm]{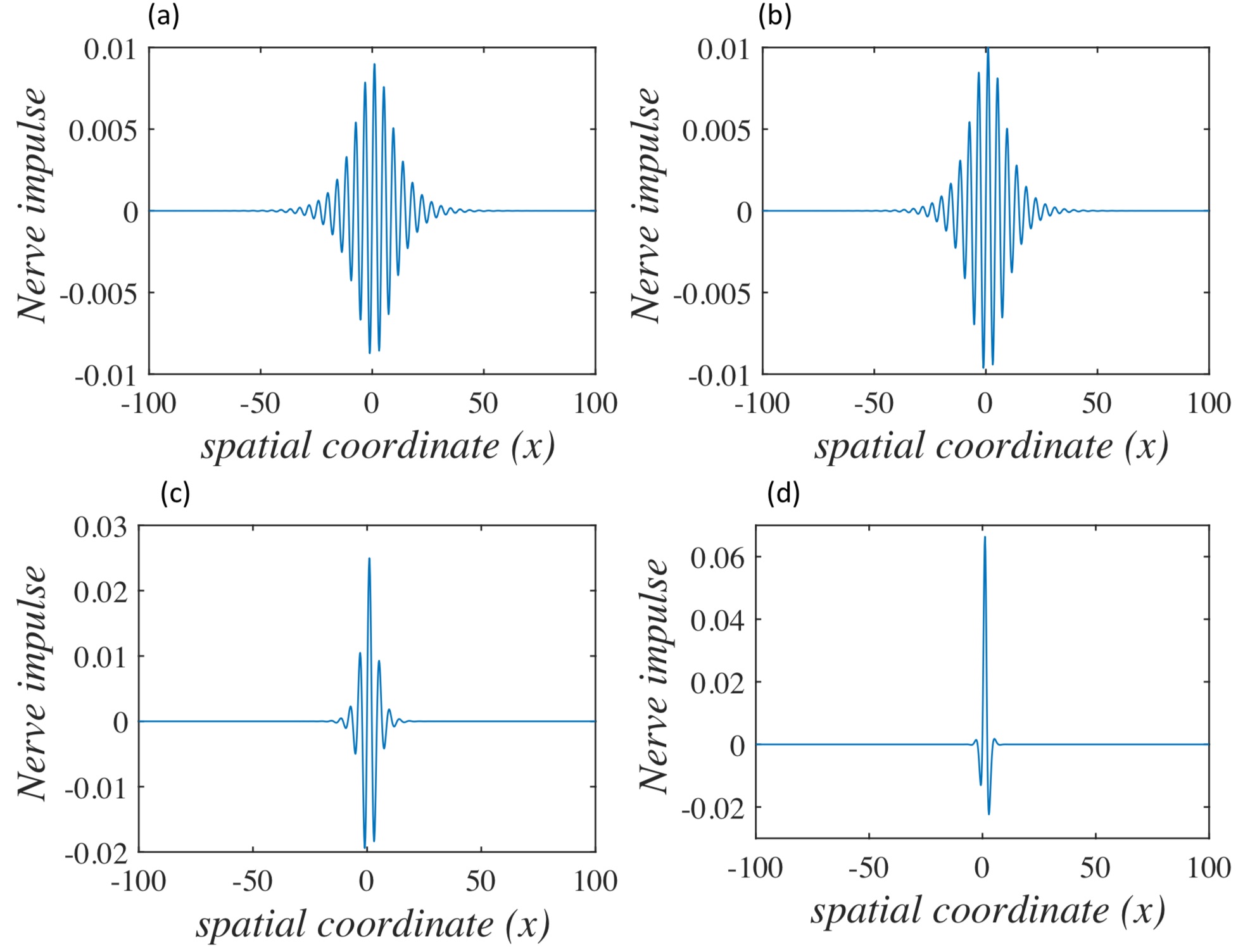}
\caption{Spatiotemporal evolution of the traveling wave profiles for the slow-fast FHR system \eqref{model}, using Eq. \eqref{Eq17}. A series of traveling pulses is shown. The multi-bump solution and disappearance of bumps with the generation of single solitary wave profiles and effects of perturbation on the impulses are presented for the parameters $\mu=0.003$, $q=1.5$, $D=0.01$, $t=10$, $A_0=0.1$ in (a) for $\epsilon=0.05$, (b) $\epsilon=0.1$, (c) $\epsilon=0.25$ and (d) $\epsilon=0.7$.}\label{fig3}
\end{figure*}

\begin{figure*}
\centering
\includegraphics[width=14cm,height=12cm]{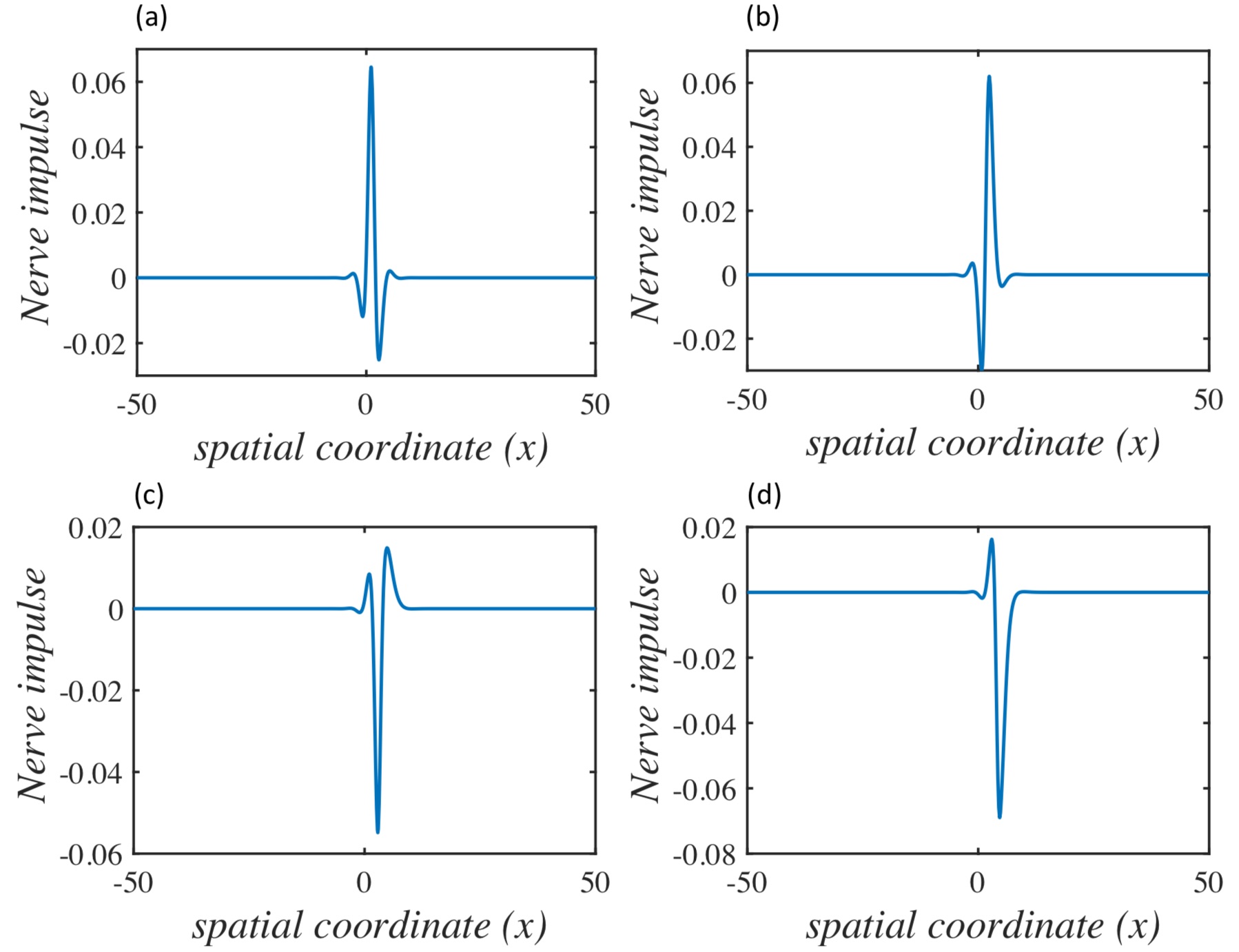}
\caption{Traveling wave profiles of analytical envelope solutions to system \eqref{model}, using Eq. \eqref{Eq17}. Effects of small parameter values $\mu$ on nerve impulses. Here $q = 1.5$, $D = 0.01$, $t = 10$, $A_0 =0.1$, $\epsilon=0.7$. Panel (a) is for $\mu = 0.000001$, panel (b) for $\mu = 0.1$, panel (c) for $\mu = 0.4$ and panel (d) for $\mu = 0.8$.}\label{fig4}
\end{figure*}

\begin{figure*}
\centering
\includegraphics[width=14cm,height=5cm]{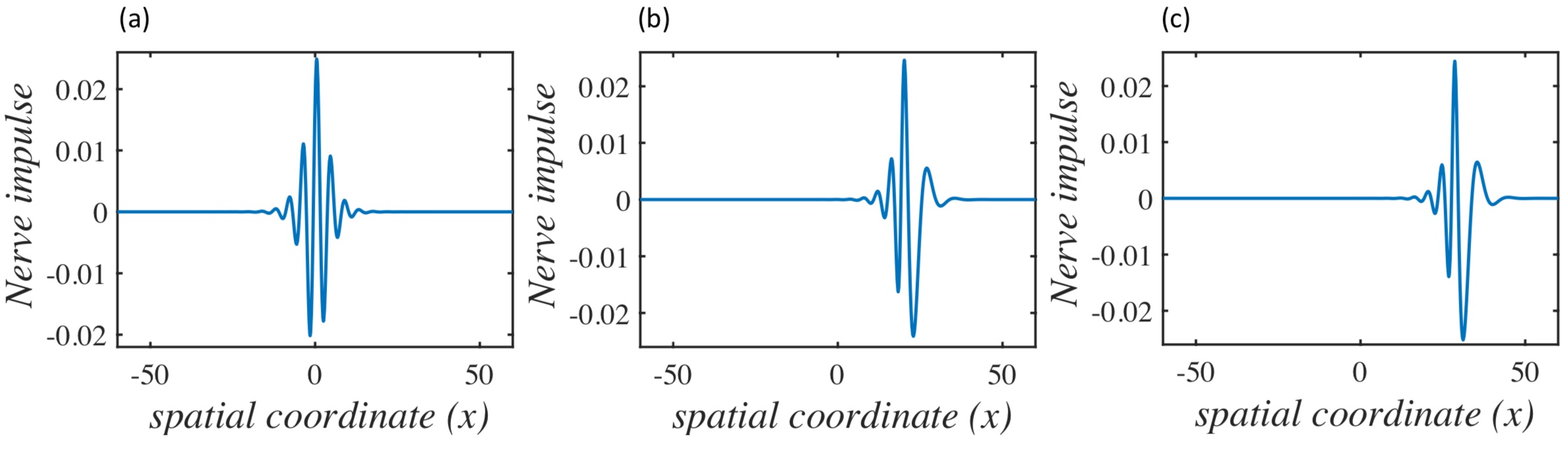}
\caption{Traveling wave profiles of spatial envelope solutions to system \eqref{model}, using Eq. \eqref{Eq17}. Effects of small diffusion coefficient $D$ on the wave profiles for the parameters $q = 1.5$, $\mu = 0.003$, $t = 10$, $A_0 =0.1$, $\epsilon=0.25$ for (a) $D = 0.001$, (b) $D = 5$ and (c) $D = 10$.}\label{fig5}
\end{figure*}

\begin{figure*}
\centering
\includegraphics[width=13cm,height=5cm]{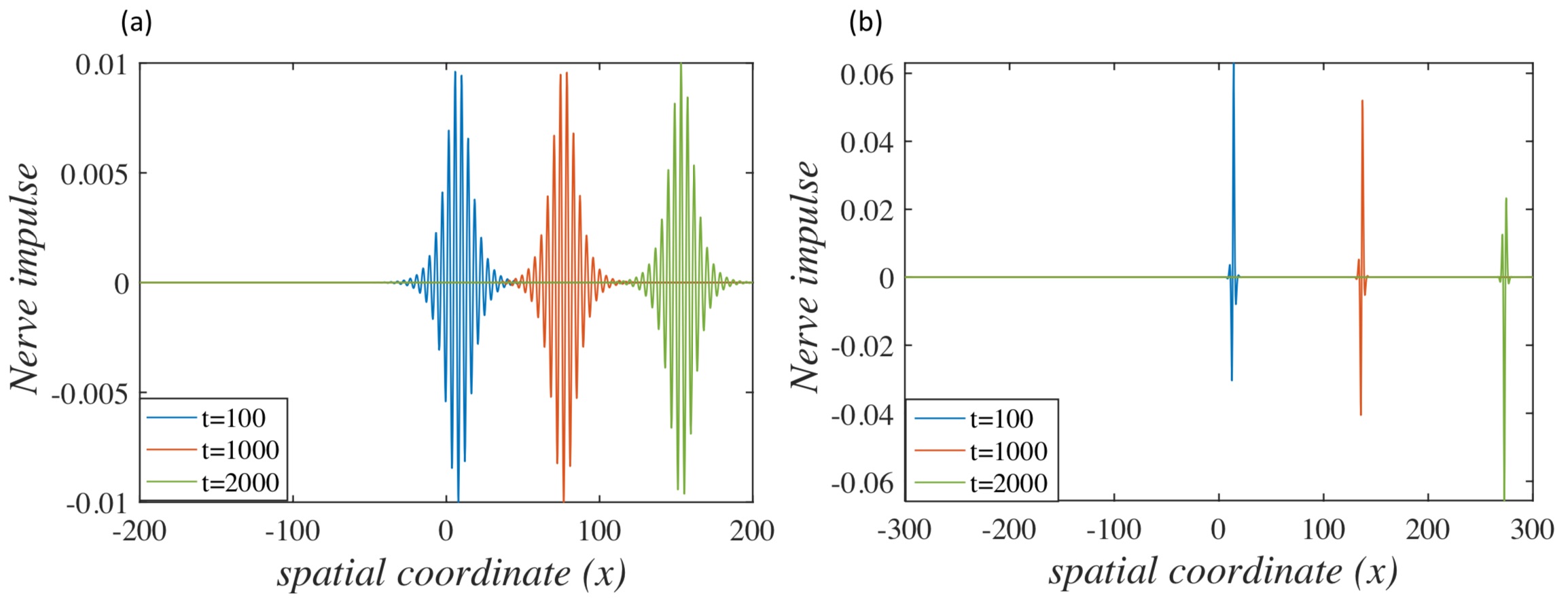}
\caption{Numerical results for traveling wave pulses at different times $t$ (shown in the insets). Multi-bump solutions in the form of (a) symmetric wave profiles with almost-same amplitudes at three different times and (b) asymmetric wave profiles with different amplitudes at the same three different times. The different colors denote the different times shown in the insets at which the traveling pulses are plotted. In both panels, we have used the parameters $\mu=0.003$, $q=1.5$, $D=0.01$, $A_0=0.1$, with $\epsilon=0.1$ in panel (a) and $\epsilon=0.7$ in panel (b).}\label{fig6}
\end{figure*}

In Fig. \ref{fig3}, we show the effects of the perturbation parameter, $\epsilon$, on the nerve impulses for fixed values of the remaining parameters. The constants $A_0$ and $q$ assume small values \cite{kakmeni2014localized}. The changes of the parameters do not affect the dynamics. The value of $q$ is considered in the region mentioned in Fig. \ref{fig1}(a). At lower perturbation $\epsilon$, we observe the envelope of a multi-bump solution for fixed diffusive coupling $D=0.01$ changing in time, where each of the bumps is unstable and transient \cite{kilpatrick2008traveling}. As the perturbation $\epsilon$ increases, there is not enough activity to generate new bumps. As a result, there is no wave propagation. Finally, at higher perturbation $\epsilon$, the nerve impulse is localized (see Fig. \ref{fig3}(d)) in the coupled network, which is an asymmetric envelope soliton. The parameter $\epsilon$ plays a major role not only in wave profiles but also in their amplitudes. It also affects the structural pattern of the wave profiles as depending on its value, it can transform the pattern into single soliton pulses.

The disappearance of bumps changes the propagation activity. The individual bump solution in the multi-bump wave profiles are transient and non-propagating. The amplitude of the impulse increases as the perturbation $\epsilon$ increases, which indicates the movement of more ions across the cell's membrane, leading to the generation of high-amplitude action potentials. This phenomenon occurs due to the fluctuations in the sequence of neuronal firing times \cite{folias2005stimulus,folias2004breathing}. However, the nature of the traveling impulses changes in time when the value of the slow-modulation parameter $\mu$ in the slow subsystem changes for fixed higher values of the perturbation parameter $\epsilon$. This represents single-soliton pulses and the amplitudes of the single pulses decrease with the change in the values of $\mu$, as shown in Fig. \ref{fig4}. The value of the slow-modulation parameter must be small in the coupled dynamics to obtain higher amplitude traveling pulses. 

Next, we study the dynamics of traveling-wave profiles for $\epsilon=0.25$ and fixed time by changing the coupling strength, $D$, shown in Fig. \ref{fig5}. They exhibit multi-pulses and the disappearance of both-end ripples changes the propagation activity of the wave profiles. We can observe how multi-pulse wave profiles emerge with surrounding standing pulses which collapses to single-traveling pulses. The solitary pulse corresponds to the envelope of a multiple bump solution. We also study the changes of the nerve impulse with respect to time $t$ in Fig. \ref{fig6}. The amplitude of the envelope of a multi-bump is independent of time (see Fig. \ref{fig6} (a)). The form and amplitude of the asymmetric envelope-soliton change with time (see Fig. \ref{fig6} (b)), therefore it is structurally unstable. The corresponding pulses are highly nonlinear envelope-solitons. 

Next, we are interested in finding the solution of Eq. \eqref{CGLE}, if the dissipation term has both real and imaginary parts (i.e., $R_i \neq 0$). Let the solution of Eq. \eqref{CGLE} be $A_{new}=Ae^{i\delta \tau_2}$. From Eq. \eqref{CGLE}, we get $\delta= - \frac{R_i}{2}$. Hence the solution is given by
\begin{equation}
A_{new}=Ae^{-\frac{iR_i}{2} \tau_2}.
\end{equation}

\section{Instability of plane waves}\label{sec5}

In the previous section, we noticed that Eq. \eqref{CGLE} has both envelope of a multi-bump and soliton wave profiles. In the following, we will find under which conditions, plane waves are stable or unstable for small perturbations. Amplitude modulated pulses emerge due to the instability of plane waves. We obtain a plane wave profile of the form
\begin{equation}
A(u_k , \tau_2) = A_0e^{i(au_k - f \tau_2)},\label{soCGLE}
\end{equation}
where $A_0$ is the plane wave amplitude and $a$ and $f$ are the wave number and angular frequency, respectively. Substituting Eq. \eqref{soCGLE} into Eq. \eqref{CGLE}, from the real part, we obtain
\begin{equation}
f=\frac{P}{2}a^2 + \frac{R_i}{2} - Q_rA_0^2\label{dp}
\end{equation}
and from the complex part,
\begin{equation}
Q_iA_0^2 + \frac{R_r}{2} = 0.
\end{equation}
Equation \eqref{dp} is known as the dispersion relation of plane waves. It is clear from Eq. \eqref{dp} that the wave angular frequency depends not only on the wave number but also on the wave amplitude.

To study the instability of the plane wave, we consider a solution of the form
\begin{equation}
A(u_k , \tau_2) = (A_0 + a_1(u_k , \tau_2))e^{i(au_k - f\tau_2 + a_2(u_k , \tau_2))},\label{soCGLEp}
\end{equation}
where the perturbation amplitude $a_1(u_k , \tau_2)$ is very small with respect to the plane wave amplitude $A_0$. Using Eqs. \eqref{CGLE}, \eqref{soCGLEp} and neglecting the nonlinear terms of the perturbations $a_1$ and $a_2$, we derive the equations
\begin{eqnarray}
-A_0{a_2}_{\tau_2} +\frac{P}{2}{a_1}_{uu} -PA_0a{a_2}_{u} +2Q_rA_0^2a_1 &=&0, \label{P1}
\end{eqnarray}
\begin{eqnarray}
{a_1}_{\tau_2} + \frac{P}{2}A_0{a_2}_{uu} +Pa{a_1}_{u} - R_ra_1 &=&0. \label{P2}
\end{eqnarray}
Equations \eqref{P1} and \eqref{P2} describe the evolution of the perturbation. We consider the solutions to Eqs. \eqref{P1} and \eqref{P2} in the form 
\begin{eqnarray}
a_1&=&a_{10}e^{i(\zeta_1 u - \zeta_2 \tau_2)}+c.c.,\label{PS1}
\end{eqnarray}
\begin{eqnarray}
a_2&=&a_{20}e^{i(\zeta_1 u - \zeta_2 \tau_2)}+c.c.,\label{PS2}
\end{eqnarray}
where $\zeta_1$ and $\zeta_2$ are the wave number of the perturbation and its corresponding propagation frequency respectively and $c.c.$ stands for complex conjugate. Generally, the wave number is real and the propagation frequency complex.

Substituting the solutions \eqref{PS1} and \eqref{PS2} in Eqs. \eqref{P1} and \eqref{P2}, we derive the following linear homogeneous equations for $a_{10}$ and $a_{20}$
\begin{eqnarray}
\left(2Q_rA_0^2 - \frac{P}{2}\zeta_1^2\right)a_{10} + iA_0(\zeta_2 - Pa\zeta_1)a_{20} &=&0,\label{PS11}\\
(-R_r-i(\zeta_2 - Pa\zeta_1))a_{10} - \frac{P}{2}A_0 \zeta_1^2a_{20} &=&0,\label{PS22}
\end{eqnarray}
which can be written in the matrix form
\begin{equation*}
LL_1=0,
\end{equation*}
where 
$L=\left( {\begin{array}{*{20}{c}}
2Q_rA_0^2 - \frac{P}{2}\zeta_1^2&iA_0(\zeta_2 - Pa\zeta_1)\\
-R_r-i(\zeta_2 - Pa\zeta_1)&-\frac{P}{2}A_0 \zeta_1^2
\end{array}} \right)$ and $L_1=\left( {\begin{array}{*{20}{c}}
{a_{10}}\\
a_{20}
\end{array}} \right)$.

Equations \eqref{PS11} and \eqref{PS22} have the nontrivial solutions, if $L$ is a singular matrix, i.e. if $\det(L)=0$, from which it results that
\begin{equation}
(\zeta_2 - Pa\zeta_1)^2=\frac{P^2\zeta_1^2}{4}\left(\zeta_1^2 - \frac{4Q_r}{P}A_0^2\right) + iR_r(\zeta_2 - Pa\zeta_1).\label{det0}
\end{equation}
If $y=\zeta_2 - Pa\zeta_1$, Eq. \eqref{det0} becomes
\begin{equation}
y^2 - iR_ry - \frac{P^2\zeta_1^2}{4}\left(\zeta_1^2 - \frac{4Q_r}{P}A_0^2\right) = 0.\label{det00}
\end{equation}
Equation \eqref{det0} is known as the dispersion relation of the perturbation. Furthermore, for a fixed value of the wave number $\zeta_1$, the term $\frac{Q_r}{P}$ plays a major role in controlling the dynamics of the angular frequency $\zeta_2$. Solving Eq. \eqref{det00}, we obtain
\begin{equation}
y= i \frac{R_r}{2} \pm \frac{\sqrt{P^2\zeta_1^2\left(\zeta_1^2 - \frac{4Q_r}{P}A_0^2\right) - R_r^2}}{2}.\label{solutiony}
\end{equation}

Depending on the sign of the discriminant, we examine the following three cases:
\begin{itemize}
\item[\bf Case 1:]{$P^2\zeta_1^2(\zeta_1^2 - \frac{4Q_r}{P}A_0^2) - R_r^2=0$, i.e., $y= i \frac{R_r}{2}$.\\The imaginary part of $\zeta_2$ is the same as the imaginary part of $y$, which is given by $\frac{R_r}{2}$. From Fig. \ref{fig7}, it is clear that $R_r<0$, which indicates the wave oscillates about its original value and after a certain period of time the perturbations die out. Thus, the plane wave is stable.
\begin{figure}
\centering
\includegraphics[width=9cm,height=5.5cm]{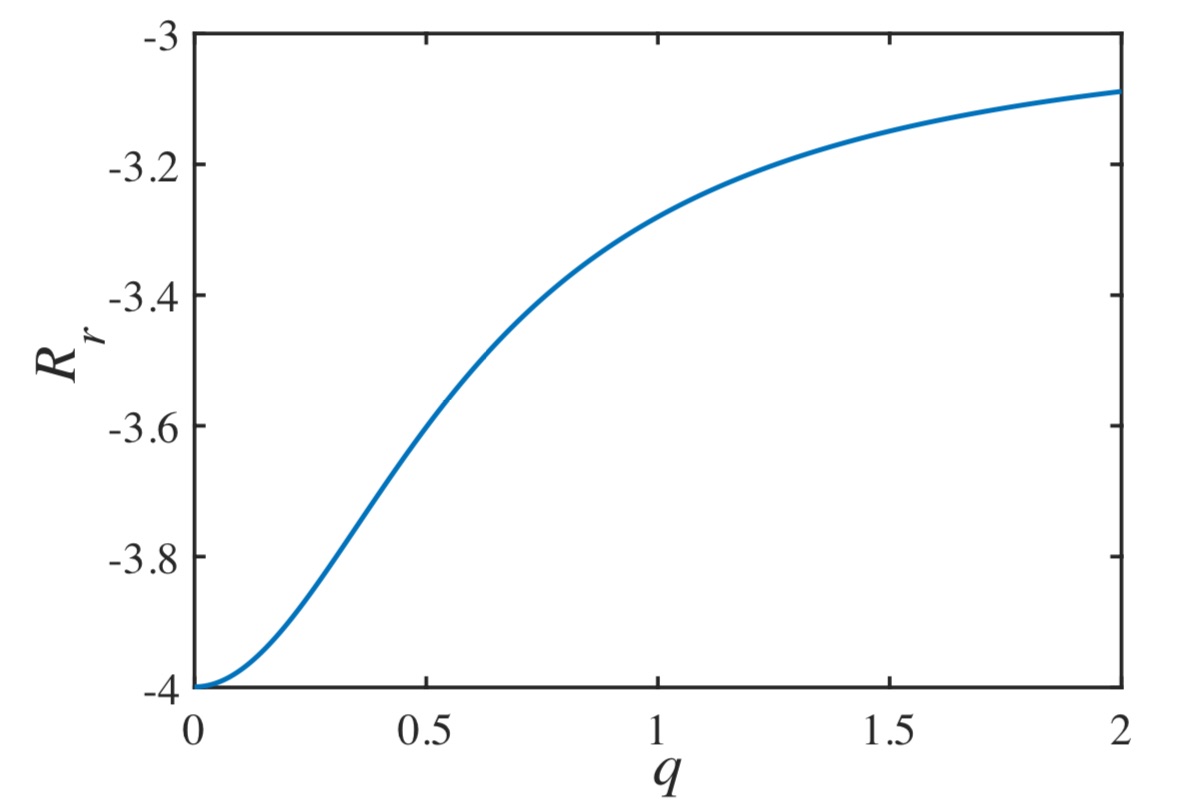}
\caption{Effects of the wave vector $q$ on the real dissipation coefficient $R_r$ for $\mu=0.003$ and $D=0.01$.}\label{fig7}
\end{figure}}

\item[\bf Case 2:]{$P^2\zeta_1^2(\zeta_1^2 - \frac{4Q_r}{P}A_0^2) - R_r^2>0$, i.e., $\Im(y)= i \frac{R_r}{2}$.\\
Similarly, the plane wave is also stable.}

\item[\bf Case 3:]{$P^2\zeta_1^2(\zeta_1^2 - \frac{4Q_r}{P}A_0^2) - R_r^2<0$, i.e., $y= i \frac{R_r}{2} \pm i\frac{\sqrt{R_r^2 - P^2\zeta_1^2(\zeta_1^2 - \frac{4Q_r}{P}A_0^2)}}{2}.$\\
As we already know the plane wave is stable if $\Im(y)<0$, which gives 
\begin{equation*}
P^2\zeta_1^2\left(\zeta_1^2 - \frac{4Q_r}{P}A_0^2\right) > 0.
\end{equation*}
It is clear from the above expression that the plane wave solution is stable if $P$ and $Q_r$ have opposite signs, i.e., if $PQ_r<0$. Similarly, we can show that the plane wave solution is unstable, i.e., the perturbations grow exponentially in time, if $PQ_r > 0$.\\
In the case of modulational instability, the local growth rate or gain is given by 
\begin{align*}
h&=\mid \Im y \mid = \mid \Im \zeta_2 \mid\\
&=\frac{1}{2}\Big(R_r + \sqrt{R_r^2 - P^2\zeta_1^2(\zeta_1^2 - \frac{4Q_r}{P}A_0^2)}\Big).
\end{align*}
The gain is maximum, i.e., the plane wave modulates itself, if the wave number $\zeta_1 = A_0 \sqrt{\frac{Q_r}{P}}$ and the corresponding gain is given by 
\begin{align}
h=\frac{1}{2}\Big(R_r + \sqrt{R_r^2 + 3Q_r^2A_0^4}\Big).\label{h}
\end{align}
It is clear from Eq. \eqref{h} that the amplitude of the plane wave and the dissipation coefficient of CGLE (Eq. \eqref{CGLE}) play an important role in controlling the growth rate of modulational instability.}
\end{itemize}

\section{Conclusions and Discussion}\label{sec6}

In this paper, we studied nonlinear excitations in a locally diffusive network of excitable slow-fast FitzHugh-Rinzel neurons. We showed that the dimension of the model can be reduced by using perturbation theory and that it can transform to an extended complex Ginzburg-Landau equation. The solitary traveling wave profiles of the voltage dynamics are also discussed. Our results show that the activities of the coupled neurons exhibit collective dynamics that cannot be observed in the activity of single neurons. Furthermore, we analyzed the propagation of wave profiles in an excitatory, diffusively coupled network of slow-fast neurons, treating it as a spatially extended medium. The model exhibits different neuro-computational features. We showed how small amplitudes and short wavelengths affect the homogeneous network and how they lead to traveling pulses. The wave speed was estimated using perturbation theory. The stable traveling pulses exist for sufficiently small $\epsilon$, however, the pulse is the envelope of the multiple traveling wave profiles in which the multiple damping terms are transient and unstable. The traveling wave profiles emerge via the appearance of multiple small-amplitude oscillations at the beginning of the wave profile.

Our analytical treatment describes envelope solitary and multiple pulses \cite{kilpatrick2008traveling} observed in numerical simulations. Successive transitions from single pulses to multi-bump solutions are described. The modulated traveling pulses and their properties are observed and are influenced by the diffusive coupling and the effects of small perturbations. Our analysis shows the appearance of multiple small waves or series of pulses in the excitable network, that are interesting to explore further as it may be possible to measure different properties of traveling pulses in excitable media. The signature of these types of time-dependent traveling-wave profiles can be effectively measured by the variations in the parameters of certain slow-fast excitable systems \cite{kakmeni2014localized,tankou2019localized}. 
Neurons can connect electrically with other nearby neurons. This is a type of weak, diffusive, coupling and is reminiscent of bursting-like activity in $\beta$-cell islets in the pancreas that secrete insulin \cite{villacorta2013wave,perez1991biophysical}. From a biological viewpoint, this shows that neurons ``communicate'' with each other and participate in the collective processing of information exchange, a portion of which is shared among them. As a result, the brain can effectively process information not only in the temporal but also in the spatial domain \cite{ermakova2009propagation,nergo1998evidence}. Neuronal ensembles can process stimuli in different or similar ways. The  waves (action potentials) exhibit nonlinear-envelope solitons having an up-and-down asymmetry in their amplitudes, as we also show in this paper. The amplitude of the impulse increases as the perturbation $\epsilon$ increases, which indicates the movement of more ions across the cell's membrane, leading to the generation of high-amplitude action potentials. This is caused by fluctuations in the sequence of neuronal firing times \cite{wig2011concepts,pinto2001spatially,raghavachari1999waves}. In extended spatial systems of neuronal populations, a transmembrane potential difference can travel across neurons by means of traveling wave propagation. Hence if one would want to measure propagation of nerve impulses in a neural network, one would need to measure the transmembrane potential difference to observe the phenomena discussed in this paper. Our work is theoretical and supported by numerical results that confirm the analytical findings. Hence, our findings are qualitative and important as they pave the way for experimentalists to look out for them in experiments.
 Theoretical and numerical results are important in understanding functional mechanisms in neural computations \cite{ma2017review,iqbal2017pattern,muruganandam1997bifurcation}. A key challenge is to analyze the intrinsic dynamics of neuronal activities and its characteristics \cite{connors1990intrinsic,izhikevich2004model,yafia2013existence}. 

\section*{Appendix A}
\label{appendix}
Equations \eqref{model2D1} and \eqref{model2D2} can be written in terms of the amplitudes $A$, $B$, $C$, $F$, $G$ and $H$ in the forms
\begin{equation}
	\begin{array}{l}
		\left(\epsilon^2 \frac{\partial^2 A}{\partial T_1^2} - \epsilon 2i\omega \frac{\partial A}{\partial T_1} - \epsilon ^2 2i\omega \frac{\partial A}{\partial T_2} - \omega ^2 A\right)e^{i\theta_k} + 
		\Big(\epsilon^2 \frac{\partial^2 \bar{A}}{\partial T_1^2} + \\ \vspace{0.2cm}
		\epsilon 2i\omega \frac{\partial \bar{A}}{\partial T_1} + \epsilon ^2 2i\omega \frac{\partial \bar{A}}{\partial T_2} - \omega ^2 \bar{A}\Big)e^{-i\theta_k}  +
		\left(-\epsilon^2 4i\omega \frac{\partial C}{\partial T_1} - \epsilon 4\omega^2 C\right)\\e^{2i\theta_k} + \left(\epsilon^2 4i\omega \frac{\partial \bar{C}}{\partial T_1} - \epsilon 4\omega ^2 \bar{C}\right)e^{-2i\theta_k}  
		+ E_0 \Big(A e^ {i\theta_k} + \overline{A} e^ {-i\theta_k} + \\ \vspace{0.2cm} \epsilon (B + C e^ {2i\theta_k} + \overline{C} e^ {-2i\theta_k})\Big) + \epsilon^2\Big(E_1 +E_2(A^2e^{2i\theta_k} + \bar{A}^2e^{-2i\theta_k} \\ \vspace{0.2cm}+ 2A\bar{A} )\Big)\left( -i\omega Ae^{i\theta_k} + i\omega \bar{A}e^{-i\theta_k}\right) 
		+\epsilon^2 \frac{E_2}{3} \Big(A^3e^{3i\theta_k} + \bar{A}^3e^{-3i\theta_k} \\ \vspace{0.2cm} + 3A^2 \bar{A}e^{i\theta_k} + 3A \bar{A} ^2 e^{-i\theta_k}\Big) + \epsilon^2 E_3 (Fe^{i \theta_k} + \bar{F}e^{-i \theta_k}) + o(\epsilon^3) = \\\vspace{0.2cm}
		D \Big((A + \epsilon \frac{\partial A}{\partial X_1} +  \epsilon ^2 \frac{\partial A}{\partial X_2} + \frac{\epsilon ^2}{2} \frac{\partial^2 A}{\partial X_1^2})e^{iq}e^{i\theta_k} + (\bar{A} + \epsilon \frac{\partial \bar{A}}{\partial X_1} +  \epsilon ^2 \frac{\partial \bar{A}}{\partial X_2} +\\\vspace{0.2cm}  \frac{\epsilon ^2}{2} \frac{\partial^2 \bar{A}}{\partial X_1^2})e^{-iq}e^{-i\theta_k} +
		\epsilon(B + \epsilon \frac{\partial B}{\partial X_1} +(C + \epsilon \frac{\partial C}{\partial X_1}) e^{2iq}e^{2i\theta_k} +  \\ \vspace{0.2cm} (\bar{C} + \epsilon \frac{\partial \bar{C}}{\partial X_1}) e^{-2iq}e^{-2i\theta_k}) -2(A e^ {i\theta_k} + \overline{A} e^ {-i\theta_k} +
		\epsilon (B + C e^ {2i\theta_k} + \\ \vspace{0.2cm} \overline{C} e^ {-2i\theta_k})) + (A - \epsilon \frac{\partial A}{\partial X_1} -  \epsilon ^2 \frac{\partial A}{\partial X_2} + \frac{\epsilon ^2}{2} \frac{\partial^2 A}{\partial X_1^2})e^{-iq}e^{i\theta_k} +\\ \vspace{0.2cm} (\bar{A} - \epsilon \frac{\partial \bar{A}}{\partial X_1} -  \epsilon ^2 \frac{\partial \bar{A}}{\partial X_2} + \frac{\epsilon ^2}{2} \frac{\partial^2 \bar{A}}{\partial X_1^2})e^{iq}e^{-i\theta_k} + \epsilon(B - \epsilon \frac{\partial B}{\partial X_1} + \\ \vspace{0.2cm} (C - \epsilon \frac{\partial C}{\partial X_1}) e^{-2iq}e^{2i\theta_k} + (\bar{C} - \epsilon \frac{\partial \bar{C}}{\partial X_1}) e^{2iq}e^{-2i\theta_k}) \Big) +  \epsilon^2 D_1 \\ \vspace{0.2cm} \Big( -i \omega A e^{iq}e^{i \theta_k} + i \omega \bar{A} e^{-iq}e^{-i \theta_k} -2( -i \omega A e^{i \theta_k} + i \omega \bar{A} e^{-i \theta_k} )  - \\ i \omega A e^{-iq}e^{i \theta_k} + i \omega \bar{A} e^{iq}e^{-i \theta_k}\Big) + o(\epsilon^3) \hspace{2.5cm} {\rm(A1)} \nonumber
		\label{app1}
	\end{array}
\end{equation}
and
\begin{equation}
	\begin{array}{l}
		\left(\epsilon \frac{\partial F}{\partial T_1} + \epsilon^2\frac{\partial F}{\partial T_2} - i\omega F\right)e^{i\theta_k} + \left(\epsilon \frac{\partial \bar{F}}{\partial T_1} + \epsilon^2\frac{\partial \bar{F}}{\partial T_2} + i\omega \bar{F}\right)e^{-i\theta_k} \\ \vspace{0.2cm} +\epsilon^2\frac{\partial G}{\partial T_1} +
		\left(\epsilon^2\frac{\partial H}{\partial T_1} - \epsilon 2i\omega H\right)e^{2i\theta_k} +
		\left(\epsilon^2\frac{\partial \bar{H}}{\partial T_1} + \epsilon 2i\omega \bar{H}\right)e^{-2i\theta_k} \\ \vspace{0.2cm} + s\Big(F e^ {i\theta_k} + \overline{F} e^ {-i\theta_k} + \epsilon (G + H e^ {2i\theta_k} + \overline{H} e^ {-2i\theta_k})\Big) - \\ E_0\Big(A e^ {i\theta_k} + \overline{A} e^ {-i\theta_k} + \epsilon (B + C e^ {2i\theta_k} + \overline{C} e^ {-2i\theta_k})\Big) + \\ o(\epsilon^3)=0. \hspace{6.3cm} {\rm(A2)} \nonumber
		\label{app2}
	\end{array}
\end{equation}

\subsubsection*{Conflict of Interest}{The authors have no conflicts to disclose.}

\subsubsection*{Author Contributions}{Arnab Mondal and Argha Mondal contributed equally to this work.}

\subsubsection*{Data Availability}{Data sharing is not applicable to this article as no new data were created or analyzed in this study.}


\begin{thebibliography}{10}
\expandafter\ifx\csname url\endcsname\relax
  \def\url#1{\texttt{#1}}\fi
\expandafter\ifx\csname urlprefix\endcsname\relax\def\urlprefix{URL }\fi
\expandafter\ifx\csname href\endcsname\relax
  \def\href#1#2{#2} \def\path#1{#1}\fi

\bibitem{mondal2018dynamics}
A.~Mondal, R.~K. Upadhyay, A.~Mondal, S.~K. Sharma, Dynamics of a modified
  excitable neuron model: {D}iffusive instabilities and traveling wave
  solutions, Chaos: An Interdisciplinary Journal of Nonlinear Science 28~(11)
  (2018) 113104.

\bibitem{izhikevich2001synchronization}
E.~M. Izhikevich, Synchronization of elliptic bursters, SIAM Review 43~(2)
  (2001) 315--344.

\bibitem{izhikevich2007dynamical}
E.~M. Izhikevich, Dynamical systems in neuroscience, MIT press, 2007.

\bibitem{ma2017review}
J.~Ma, J.~Tang, A review for dynamics in neuron and neuronal network, Nonlinear
  Dynamics 89~(3) (2017) 1569--1578.

\bibitem{iqbal2017pattern}
N.~Iqbal, R.~Wu, B.~Liu, Pattern formation by super-diffusion in
  {F}itz{H}ugh-{N}agumo model, Applied Mathematics and Computation 313 (2017)
  245--258.

\bibitem{muruganandam1997bifurcation}
P.~Muruganandam, M.~Lakshmanan, Bifurcation analysis of the travelling waveform
  of {F}itz{H}ugh-{N}agumo nerve conduction model equation, Chaos: An
  Interdisciplinary Journal of Nonlinear Science 7~(3) (1997) 476--487.

\bibitem{corson2009asymptotic}
N.~Corson, M.~A. Aziz-Alaoui, Asymptotic dynamics of {H}indmarsh-{R}ose
  neuronal system, Dynamics of Continuous, Discrete and Impulsive Systems,
  Series B: Applications and Algorithms~(16) (2009) 535.

\bibitem{fitzhugh1961impulses}
R.~FitzHugh, Impulses and physiological states in theoretical models of nerve
  membrane, Biophysical Journal 1~(6) (1961) 445--466.

\bibitem{rinzel1986nonlinear}
J.~Rinzel, Y.~S. Lee, Nonlinear {O}scillations in {B}iology and {C}hemistry
  (1986).

\bibitem{rinzel1987mathematical}
J.~Rinzel, Mathematical topics in population biology, morphogenesis, and
  neurosciences, Lecture Notes in Biomathematics 71 (1987) 267--281.

\bibitem{townsend2018detection}
R.~G. Townsend, P.~Gong, Detection and analysis of spatiotemporal patterns in
  brain activity, PLoS Computational Biology 14~(12) (2018) e1006643.

\bibitem{milton1993spiral}
J.~G. Milton, P.~H. Chu, J.~D. Cowan, Spiral waves in integrate-and-fire neural
  networks, Advances in neural information processing systems (1993)
  1001--1001.

\bibitem{kondo2010reaction}
S.~Kondo, T.~Miura, Reaction-diffusion model as a framework for understanding
  biological pattern formation, Science 329~(5999) (2010) 1616--1620.

\bibitem{meier2015bursting}
S.~R. Meier, J.~L. Lancaster, J.~M. Starobin, Bursting regimes in a
  reaction-diffusion system with action potential-dependent equilibrium, PloS
  One 10~(3) (2015) e0122401.

\bibitem{feng2016spike}
P.~Feng, J.~Zhang, W.~Wang, Spike-like solitary waves in incompressible
  boundary layers driven by a travelling wave, Chaos: An Interdisciplinary
  Journal of Nonlinear Science 26~(6) (2016) 063104.

\bibitem{belykh2005synchronization}
I.~Belykh, E.~De~Lange, M.~Hasler, Synchronization of bursting neurons: {W}hat
  matters in the network topology, Physical Review Letters 94~(18) (2005)
  188101.

\bibitem{connors1993generation}
B.~W. Connors, Y.~Amitai, Generation of epileptiform discharges by local
  circuits in neocortex, Epilepsy: Models, Mechanisms and Concepts (1993)
  388--424.

\bibitem{lance1993current}
J.~W. Lance, Current concepts of migraine pathogenesis, Neurology 43~(6 Suppl
  3) (1993) S11--5.

\bibitem{connors1990intrinsic}
B.~W. Connors, M.~J. Gutnick, Intrinsic firing patterns of diverse neocortical
  neurons, Trends in Neurosciences 13~(3) (1990) 99--104.

\bibitem{izhikevich2004model}
E.~M. Izhikevich, Which model to use for cortical spiking neurons?, IEEE
  Transactions on Neural Networks 15~(5) (2004) 1063--1070.

\bibitem{yafia2013existence}
R.~Yafia, M.~A. Aziz-Alaoui, Existence of periodic travelling waves solutions
  in predator prey model with diffusion, Applied Mathematical Modelling 37~(6)
  (2013) 3635--3644.

\bibitem{wig2011concepts}
G.~S. Wig, B.~L. Schlaggar, S.~E. Petersen, Concepts and principles in the
  analysis of brain networks, Annals of the New York Academy of Sciences
  1224~(1) (2011) 126--146.

\bibitem{folias2005stimulus}
S.~E. Folias, P.~C. Bressloff, Stimulus-locked traveling waves and breathers in
  an excitatory neural network, SIAM Journal on Applied Mathematics 65~(6)
  (2005) 2067--2092.

\bibitem{kilpatrick2010effects}
Z.~P. Kilpatrick, P.~C. Bressloff, Effects of synaptic depression and
  adaptation on spatiotemporal dynamics of an excitatory neuronal network,
  Physica D: Nonlinear Phenomena 239~(9) (2010) 547--560.

\bibitem{pinto2001spatially}
D.~J. Pinto, G.~B. Ermentrout, Spatially structured activity in synaptically
  coupled neuronal networks: I. {T}raveling fronts and pulses, SIAM Journal on
  Applied Mathematics 62~(1) (2001) 206--225.

\bibitem{kilpatrick2008traveling}
Z.~P. Kilpatrick, S.~E. Folias, P.~C. Bressloff, Traveling pulses and wave
  propagation failure in inhomogeneous neural media, SIAM Journal on Applied
  Dynamical Systems 7~(1) (2008) 161--185.

\bibitem{ermakova2009propagation}
E.~A. Ermakova, E.~E. Shnol, M.~A. Panteleev, A.~A. Butylin, V.~Volpert, F.~I.
  Ataullakhanov, On propagation of excitation waves in moving media: {T}he
  {F}itz{H}ugh-{N}agumo model, PloS One 4~(2) (2009) e4454.

\bibitem{villacorta2013wave}
J.~A. Villacorta-Atienza, V.~A. Makarov, Wave-processing of long-scale
  information by neuronal chains, PloS One 8~(2) (2013) e57440.

\bibitem{kakmeni2014localized}
F.~M. Kakmeni~Moukam, E.~M. Inack, E.~M. Yamakou, Localized nonlinear
  excitations in diffusive {H}indmarsh-{R}ose neural networks, Physical Review
  E 89~(5) (2014) 052919.

\bibitem{tankou2019localized}
A.~S. Tankou~Tagne, C.~N. Takembo, H.~G. Ben-Bolie, P.~Owona~Ateba, Localized
  nonlinear excitations in diffusive memristor-based neuronal networks, PloS
  One 14~(6) (2019) e0214989.

\bibitem{malfliet1996tanh}
W.~Malfliet, W.~Hereman, The tanh method: I. {E}xact solutions of nonlinear
  evolution and wave equations, Physica Scripta 54~(6) (1996) 563.

\bibitem{wazwaz2005tanh}
A.-M. Wazwaz, The tanh method: {E}xact solutions of the sine-{G}ordon and the
  sinh-{G}ordon equations, Applied Mathematics and Computation 167~(2) (2005)
  1196--1210.

\bibitem{stefanescu2008low}
R.~A. Stefanescu, V.~K. Jirsa, A low dimensional description of globally
  coupled heterogeneous neural networks of excitatory and inhibitory neurons,
  PLoS Computational Biology 4~(11) (2008) e1000219.

\bibitem{rinzel1982bursting}
J.~Rinzel, W.~C. Troy, Bursting phenomena in a simplified {O}regonator flow
  system model, The Journal of Chemical Physics 76~(4) (1982) 1775--1789.

\bibitem{wojcik2015voltage}
J.~Wojcik, A.~Shilnikov, Voltage interval mappings for an elliptic bursting
  model, in: Nonlinear Dynamics New Directions, Springer International
  Publishing Switzerland, 2015, pp. 195--213.

\bibitem{mondal2019firing}
A.~Mondal, S.~K. Sharma, R.~K. Upadhyay, A.~Mondal, Firing activities of a
  fractional-order {F}itz{H}ugh-{R}inzel bursting neuron model and its coupled
  dynamics, Scientific Reports 9~(1) (2019) 1--11.

\bibitem{nergo1998evidence}
C.~A. Del~Negro, C.~F. Hsiao, S.~H. Chandler, A.~Garfinkel, Evidence for a
  novel bursting mechanism in rodent trigeminal neurons, Biophysical Journal
  75~(1) (1998) 174--182.

\bibitem{mondal2021spatiotemporal}
A.~Mondal, A.~Mondal, S.~Kumar~Sharma, R.~Kumar~Upadhyay, C.~G. Antonopoulos,
  Spatiotemporal characteristics in systems of diffusively coupled excitable
  slow-fast {FitzHugh-Rinzel} dynamical neurons, Chaos: An Interdisciplinary
  Journal of Nonlinear Science 31~(10) (2021) 103122.

\bibitem{perez1991biophysical}
M.~Perez-Armendariz, C.~Roy, D.~C. Spray, M.~V. Bennett, Biophysical properties
  of gap junctions between freshly dispersed pairs of mouse pancreatic beta
  cells, Biophysical Journal 59~(1) (1991) 76--92.

\bibitem{raghavachari1999waves}
S.~Raghavachari, J.~A. Glazier, Waves in diffusively coupled bursting cells,
  Physical Review Letters 82~(14) (1999) 2991.

\bibitem{folias2004breathing}
S.~E. Folias, P.~C. Bressloff, Breathing pulses in an excitatory neural
  network, SIAM Journal on Applied Dynamical Systems 3~(3) (2004) 378--407.

\end{thebibliography}

\end{document}